\documentclass[preprint,prc,aps,showpacs]{revtex4-1}
\usepackage{graphicx,epsfig} 

\begin{document}

\title{A non-equilibrium picture of the chemical freeze-out in hadronic thermal models}

\author{Leonardo P. G. De Assis}\email{lpgassis@cbpf.br}
\affiliation{Centro Brasileiro de Pesquisas F\'isicas\\
Rua Dr. Xavier Sigaud 150, 22290-180 Rio de Janeiro--RJ, Brasil}

\author{S\'{e}rgio B. Duarte}
\affiliation{Centro Brasileiro de Pesquisas F\'isicas\\
Rua Dr. Xavier Sigaud 150, 22290-180 Rio de Janeiro--RJ, Brasil}    

\author{Marcelo Chiapparini}
\affiliation{Instituto de
F\'{\i}sica, Universidade do Estado do Rio de Janeiro\\
Rua S\~{a}o Francisco Xavier 524, 20550-900 Rio de Janeiro--RJ, Brasil}        
 
\author{ Luciana R. Hirsch}
\affiliation{Departamento de F\'{\i}sica, Universidade Tecnol\'{o}gica Federal do Paran\'{a}\\
Av. Sete de Setembro, 3165, 80230-901 Curitiba--PR, Brasil}        
        
\author{Antonio Delfino}
\affiliation{Instituto de F\'isica, Universidade Federal Fluminense\\
Av. Gal. Milton Tavares de Souza, 24210-346 Niter\'oi--RJ, Brasil}
                
\begin{abstract}
Thermal models have proven to be an useful and simple tool used to make theoretical predictions and data analysis in relativistic and ultra-relativistic heavy ion collisions. A new version of these models is presented here, incorporating a non equilibrium feature to the description of the intermediate fireball state formed at the chemical freeze-out. Two different effective temperatures are attributed to the expanding fireball, regarding its baryonic and mesonic sectors. The proposal is not merely to include an additional degree of freedom to reach a better adjustment to the data, but to open a room in the model conception for considerations on the non-equilibrium scenario of the system evolution.  A set of well consolidated data for particles production is used to validated the reformulated version of thermal models presented here. A rather good performance of the extended version was verified, both for the quality of particle ratio data fittings as well as for describing the asymptotic energy behavior of temperatures and baryochemical potential of the colliding nuclear system.

\end{abstract}

\pacs{25.75.-q,13.85.-t,13.75.-n}

\maketitle

\section{Introduction}

A significant amount of experimental data for particle production in high energy heavy ion collisions (10 - 200 GeV/$A$ at center of mass) has been accumulated during the last two decades. Many different theoretical attempts have tried to described these data using thermal models in the approximation of global thermal equilibrium, considering  just one freeze-out temperature \cite{kraus2009a}-\cite{hirsh2010b}. Because of its simplicity, these models are widely used to treat and explain the data. However, thermal models often are not able to describe adequately all multiplicities of hadrons \cite{cleymans1994r}. For instance, the abundance of strange particles are overestimate and the pion yields are underestimated. In addition to that, particle number density and total energy density calculated in chemical freezing are overestimated as well \cite{yen1999r}. To overcome these limitations some features have been incorporated into the model, like the inclusion of a phenomenological parameter $\gamma_{s}$ to adjust the strangeness production in the reaction process \cite{rafelski1991r}-\cite{becattini1998r}, or even the inclusion of effects of excluded volume \cite{rischke1991r}-\cite{braum1999}. However, the main criticism is that the oversimplified conception of the model makes its connection with more elaborated models \cite{bass2000}-\cite{huovinen2008} a hard task. This happens for example when trying to compare results from thermal models with those from hydrodynamical models, non equilibrium kinetic equation approach or even with others phenomenological statistical model including the formation of separated thermal subsystem \cite{becattini2009}. Non-equilibrium proposals for a phenomenological treatment of the hadronic system has been recalled with the advent of lattice calculation results, showing that the hadronization critical temperature differs significantly from the chemical freeze-out temperature estimated, using thermal conventional models, by tens of MeV \cite{noronha11}-\cite{noronha13}. The inclusion of non-equilibrium features in handling with hadron production in high energy collision were revived trying to justify the difference between critical and freeze out temperatures in these calculations. However, the used arguments were not strong enough to be compatible with RHIC results few years latter \cite{noronha14}-\cite{noronha15}. The solution is still far from being  conclusive, but some effort have been made more recently in this direction \cite{noronha}. The main idea goes back to the Hagedorn's states, introduced in the 1960's, to quickly drive hadrons to chemical equilibrium \cite{noronha14}. The introduced gas of hadronic resonant states should have a mass spectrum parametrized by the Hagedorn's exponential form using a critical temperature, the Hagedorn temperature. The discussion of the possibility of have different values of Hagedorn temperature for baryons and mesons in the hadronization process was addressed in Ref.\cite{plb2004}. However, the inclusion of the highest mass branch of observed hadronic resonant states in the Hagedorn exponential parametrization of the spectrum fails, and the critical temperature parameter value can not be well defined \cite{lu2002b}-\cite{choi2011}. 

In the present work we assume that the chemical freeze-out stage is not characterized by a global thermal equilibrium with a single temperature of the expanding fireball. Instead of this, it was introduced a picture of the  fireball freeze-out stage out of equilibrium having baryonic and mesonic sectors characterised by two distinct temperatures, which could be considered as being an effective thermodynamic quantities associated to these sectors. Indeed, we have no well founded reason to support the conventional thermal model statement of equal thermal evolution for mesons and baryons during the whole evolution of the system. The chemical freeze-out can occur in non-equilibrium condition before the kinetic freeze-out regime. Only after the later stage it is believed that the thermodynamic equilibrium is reached. Consequently, only simplicity arguments supports the isothermal fireball assumption for the chemical freeze-out. 

Since the pioneering works by Fermi \cite{Fermi50} and Heisenberg \cite{Heis49} on thermal models, the idea of treating baryons and mesons as distinct thermal sectors was suggested. In these works baryons arise merged in a meson cloud and the description for particle multiplicity are essentially statistical in nature. When the concept of temperature is introduced in Fermi's work, it is defined by relating the number of particles with the energy density of the system by mean of the Stefans-Boltzmann law. The Hagedorn resonance mass spectra mentioned previously is nothing more than another way to associate an effective temperature to the thermal state of the system. In this case the assumption is that baryons and mesons are produced in thermal contact with a correspondent Hagedorn thermal  reservoir, defined by the observable mass spectra of baryons and mesons. Different effective temperature values for mesons and baryons are extracted from this association \cite{lu2002b}-\cite{choi2011}. In addition, due to the experimental difficulties in establish the observable particle resonance spectra, some ambiguities values of these effective temperatures can be found \cite{choi2011}.

In our case, we are assuming that baryons and mesons are directly in contact with the thermal vacuum from where particles are produced. This is an image compatible with the use of a grand canonical ensemble behind the Fermi distribution (for baryons) and Bose-Einstein (for meson) employed to determine the density of particles at the freeze-out in thermal models. At the beginning of the chemical freeze out mesons and baryons are produced from an initial  thermal vacuum state, whereas at the final stage, and due to the depletion of energy density of the expanding system, the  vacuum should be considered at a different thermal state. The meson freeze-out is associated to this  final stage. We remark that these two effective temperatures introduced here in the thermal model still  preserves the relatively simple model skill to perform calculation, bringing  this kind of approach closer to a more detailed description of the fireball expansion. Having in mind this qualitative arguments, a  necessary condition to validate this alternative model is the improving of the results obtained with the conventional calculation. Thus, as a preliminary task, we applied the two-temperatures model to different collisional processes with well established data set for particle ratios, and comparing results with those obtained with the conventional model. In addition, we discuss the behavior of freeze out temperatures, entropy and baryonic chemical potential for increasing collision energy of nucleus-nucleus reaction.    

This work is organized presenting firstly in Section II, the main changes introduced in the conventional thermal model ($1T$-Model) due to the two-temperatures description ($2T$-Model). The results obtained for particle ratios within the modified model and they comparison with the results of a conventional thermal calculation is shown in Section III. The analysis of temperature, entropy and chemical potential behavior with the increasing collision energy is presented in Section IV. And finally, our conclusions and final discussions are given in Section V.

\section{Thermal model with two freeze-out temperatures}

In thermal models the particle density of the $k$th particle is obtained using the distribution functions of baryons and mesons given by the
Fermi-Dirac and Bose-Einstein distributions functions respectively, in the following way 
\begin{equation}
n_{k}=\frac{\gamma_{k}}{2\pi^{2}}\int\frac{p^{2}}{\exp\left[  \left(
\epsilon_{k}-\mu_{k}\right)  /T_{i}\right]  \pm1}dp,\label{eq1}
\end{equation}
where $\epsilon_{k}=\sqrt{p^{2}+m_{k}^{{2}}}$, the minus (plus) sign 
stands for the Fermi-Dirac (Bose-Einstein) distribution, $\gamma
_{k}$ is the spin degeneracy and $\mu_{k}$ is the chemical
potential, using $T_i\;(i=1,2$; baryon,meson) for temperature of different sectors. The chemical potential of the $k$th particle $\mu_k$ is calculated using the global chemical potentials $\{\mu_B,\mu_{{I3}},\mu_S\}$ as $\mu_k=B_k\mu_B+I3_{k}\mu_{I3}+S_k\mu_S$, where $\{B_k,I3_{k},S_k\}$ are the baryonic, isospin and strangeness numbers of the particle respectively. The global chemical potentials are constrained to the conservation of global baryonic, isospin and strangeness charges as
\begin{eqnarray}
V\sum_k B_k n_k &=&Z_1+Z_2+N_1+N_2 \label{qb}\\
V\sum_k I3_{k} n_k &=& \frac{1}{2}(Z_1+Z_2-N_1-N_2) \label{qi3}\\
  \sum_k S_k n_k &=& 0 , \label{qs}
\end{eqnarray}
where $Z_i$ and $N_i$ ($i=1,2$; target, projectile) are the number of protons and neutrons of the colliding nuclei and $V$ is the volume of the fireball. This procedure can be understood in connection with the Generalized Gibbs Ensemble prescription \cite{gge}.

The complete lightest baryonic octet, decuplet, and mesonic nonet are included, together with around two hundred of other baryonic and mesonic resonances with masses up to 2 GeV.  It is also important to point out that the particle population calculated by Eqs.~(\ref{eq1}-\ref{qs})  (to be compared to the data) should be modified due to the decay chain of resonances, a process known as the feed-down of particle populations. We reformulated the resonance decay feeding processes of the conventional thermal model in order to make it consistent with the two-temperatures approach. For that, the final density of produced particles are given by,

\begin{equation}
n_k^f=n_k+\sum_j \:\Gamma_{j\rightarrow k} \:\: n_j^{}, \label{feed}
\end{equation}
where $n_k$ is the thermal particle density calculated using Eq.~(\ref{eq1}), and $\Gamma_{j\rightarrow k}$ is the probability of particle $j$ to decay into particle $k$,  taken from experimental particles decay data. Populations of thermal particle $n_k$ are determined consistently with Eqs.~(\ref{qb})-(\ref{qs} using in Eq.~(\ref{eq1}) the specific effective temperature for baryons and mesons. 

\section{Particles ratios results in $2T$-Model}

The results obtained from the conventional
model with one temperature, referred in this work as $1T$-Model is now compared with those from the two-temperatures model, the introduced $2T$-Model. We have performed 
the conventional calculation by using the $2T$-Model code with the constraint $\ T_{b}=T_{m}$. The two-temperatures code was written from a previous version of thermal model developed by two of the authors \cite{hirsh2010}-\cite{hirsh2010b}. A Monte Carlo sampling of the model parameter values covering a wide range of the domain was taken in order to speed up the computational calculation. Results of the parameter values (temperatures and baryochemical potentia) for the best fitting of the particle population ratios produced in  the five systems under analysis are compared in Tables \ref{table1} (1T-Model) and \ref{table2} (2-Tmodel). The $C1$ data set was extracted from Ref. \cite{Cleymans1999}, $C2$ data was taken from Ref. \cite{kraus2007} and $C3$, $C4$ and $C5$ were taken from \cite{Abelev}. To compare the quality of the fittings we also show the correspondent $\chi_{d.o.f}^{2}$ values of the data adjustment. 

\begin{table}[h,t]
\caption{\label{table1}Results for the thermodynamic parameters for different collisions using conventional $1T$-Model. The error bars in the  temperature and baryonic chemical potential are established as the isolated variation in these quantities which leads to a 10\% of change in the optimal $\chi_{d.o.f}^{2}$ value.}
\begin{tabular}{|c|c|c|c|c|c|}\hline\hline
Sets & $T\left(  MeV\right)  $ & $\mu_{b}\left(  MeV\right)  $ &
$\chi_{d.o.f}^{2}$ & $\sqrt{s_{NN}} \left(  GeV\right)$ & $Ions$\\\hline
$C1$ & $29.6{\LARGE}_{(0.05)}^{(0.63)}$ & $772.9{\LARGE}_{(0.25)}^{(7.25)}$ & $0.160$ & $2.32$ & $Ni+Ni$\\\hline
$C2$ & $147.4{\LARGE}_{(2.6)}^{(4.35)}$ & $287.2{\LARGE}_{(16.4)}^{(14.9)}$ & $1.667$ & $17.3$ & $Pb+Pb$\\\hline
$C3$ & $150.5{\LARGE}_{(8.70)}^{(9.47)}$ & $60.5{\LARGE}_{(7.67)}^{(8.51)}$ & $7.435$ & $62.4$ & $Au+Au$\\\hline
$C4$ & $156.50{\LARGE}_{(7.67)}^{(8.55)}$ & $29.50{\LARGE}_{(6.74)}^{(6.89)}$ & $ 5.876$ & $130$ & $Au+Au$\\\hline
$C5$ & $169.50{\LARGE}_{(7.42)}^{(7.88)}$ & $23.00{\LARGE}_{(6.80)}^{(7.51)}$ & $ 3.159$ & $200$ & $Au+Au$\\ \hline\hline
\end{tabular}
\end{table}

\begin{table}[h,t]
\caption{\label{table2}Thermodynamic parameters for different collisions obtained by using the $2T$-Model. The error bars are defined as previously in Table \ref{table1}.}
\begin{tabular}{|c|c|c|c|c|c|c|}\hline\hline
Sets & $T_{m}\left(  MeV\right)  $ & $T_{b}\left(
MeV\right)  $ & $\mu_{b}\left(  MeV\right)  $ & $\chi_{d.o.f}^{2}$ &
$\sqrt{s_{NN}} \left(  GeV\right)$ & $Ions$\\\hline
$C1$ & $27.9{\LARGE}_{(0.09)}^{(2.19)}$ & $29.6{\LARGE}_{(0.48)}^{(0.24)}$ & $772.0{\LARGE }_{(10.35)}^{(1.0)}$ & $0.9666$ & $2.32$ & $Ni+Ni$\\\hline
$C2$ & $133.36{\LARGE}_{(0.44)}^{(1.96)}$ & $140.05{\LARGE}_{(1.85)}^{(0.20)}$ & $251.80{\Large}_{(12.80)}^{(6.60)}$ & $0.9997$ & $17.3$ & $Pb+Pb$\\\hline
$C3$ & $131.50{\LARGE}_{(0.87)}^{(1.78)}$ & $145.50{\LARGE}_{(1.38)}^{(0.66)}$ & $59.0{\LARGE }_{(6.35)}^{(0.87)}$ & $0.9996$ & $62.4$ & $Au+Au$\\\hline
$C4$ & $129.74{\LARGE}_{(0.19)}^{(7.24)}$ & $146.32{\LARGE}_{(4.08)}^{(0.17)}$ & $24.91{\LARGE }_{(0.16)}^{(2.24)}$ & $0.9999$ & $130$ & $Au+Au$\\\hline
$C5$ & $128.01{\LARGE}_{(0.09)}^{(2.19)}$ & $147.36{\LARGE}_{(0.48)}^{(0.24)}$ & $23.21{\LARGE }_{(9.41)}^{(0.89)}$ & $1.0$ & $200$ & $Au+Au$\\ \hline\hline
\end{tabular}
\end{table}

Comparing the results we see that in all cases both effective temperatures in Table \ref{table2} ($T_{b}$ and $T_{m}$) are lower than the single freeze out temperature in Table \ref{table1} ($T$). Another direct observation is that the fits in Table \ref{table2} present a better quality when compared with those in Table \ref{table1}, which is reflected in the $\chi_{d.o.f}^{2}$ values.

\begin{figure}[h]
	\includegraphics[width=165mm]{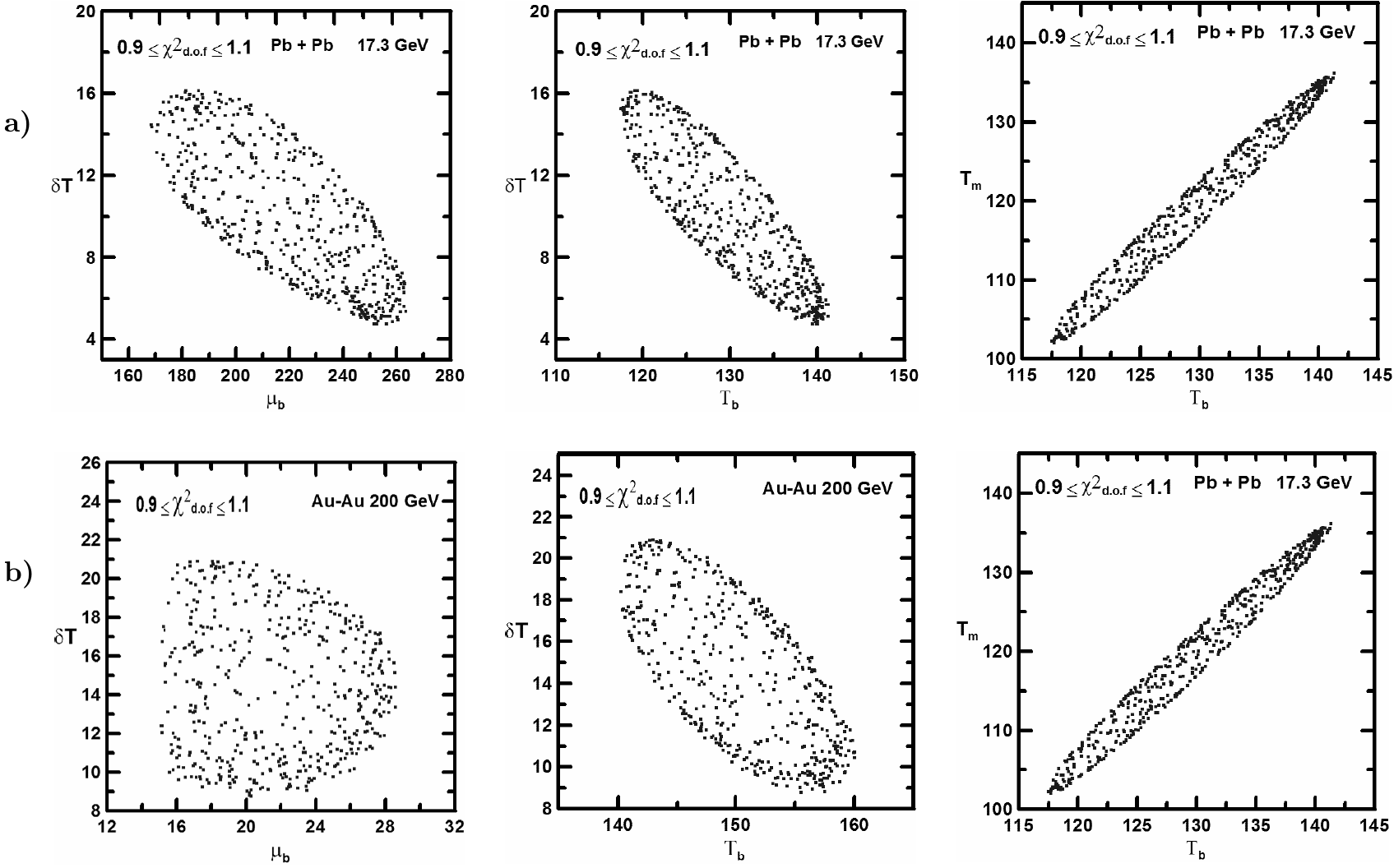}
	\caption{\label{fig1}Plots for thermal parameters of Pb + Pb collisions at  $\sqrt{s_{NN}}=17.3$ GeV in panel (a). In panel (b) we have the same plots for Au + Au at  $\sqrt{s_{NN}}=200.0$ GeV. In both calculation it was requested that the parameters values leads to a particle data rate fitting with $0.9 <\chi_{d.o.f}^{2}<1.1$. In the left column we have $\delta T\; vs\;\mu_b$ plots, in the central column we have $\delta T\;vs \;T_b$ plots and in the right column we show the points in the ($T_{b} , T_{m}$) plane.}
\end{figure}

In Figure \ref{fig1} we show plots of the parameter values for data adjustment with $chi_{d.o.f}^{2}$ in a narrow range ( $ 0.9 < \chi_{d.o.f}^{2} < 1.1 $) for the systems Pb+Pb at $17.3$ GeV and Au+Au at $200$ GeV. Each point corresponds to values of $\{T_{b}, T_{m}, \mu_{b}\}$ in the parametric space. In left panels the difference $\delta T = T_{b}-T_{m}$ are plotted against the baryonic chemical potential. A set of five hundred points are generated covering the $\chi_{d.o.f}^{2}$ window. A Monte Carlo sort of the parameters values was performed exploring a wide domain of the parameters in order to speed up the computational calculation. In central panels we depicted the same difference against the baryonic effective temperature, and in right panel we have the point spread out in the ($T_{b} , T_{m}$) plane. Similar plots were obtained for the other analyzed system (not shown) with the equivalent density of points in this narrow $\chi_{d.o.f}^{2}$ window. The values of $\delta T$ in these plots does not reach null values (($T_{b}=T_{m}=T$), indicating that $1T$-Model is not able reproduce data when the requested quality of adjustment is established. 
Except for the lightest system with lowest collision energy it was possible to have the data adjustment with $\chi_{d.o.f}^{2}$ in the established range with both models.

\begin{figure}[h]
	\includegraphics[width=140mm]{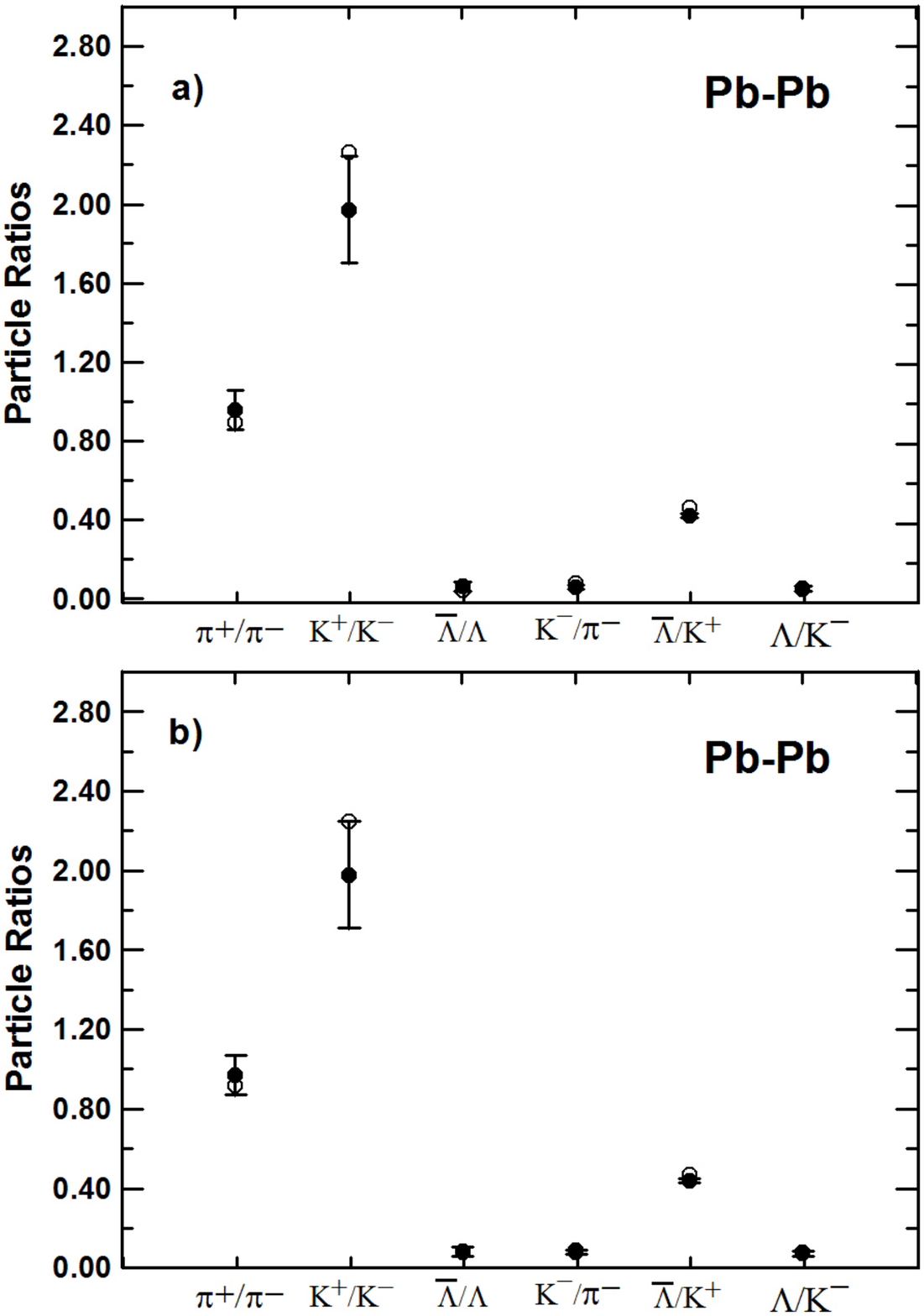}
	\caption{\label{fig2}Plots for particle ratios of Pb + Pb  at $\sqrt{s_{NN}}$=$17.3$ GeV fitted using the $1T$-Model, plot (a), and the $2T$-Model, plot (b), both with open circles. The full circles with error bars are the data taken from Refs.\cite{Cleymans1999} and \cite{kraus2007}. }
\end{figure}

\begin{figure}[h]
	\includegraphics[width=140mm]{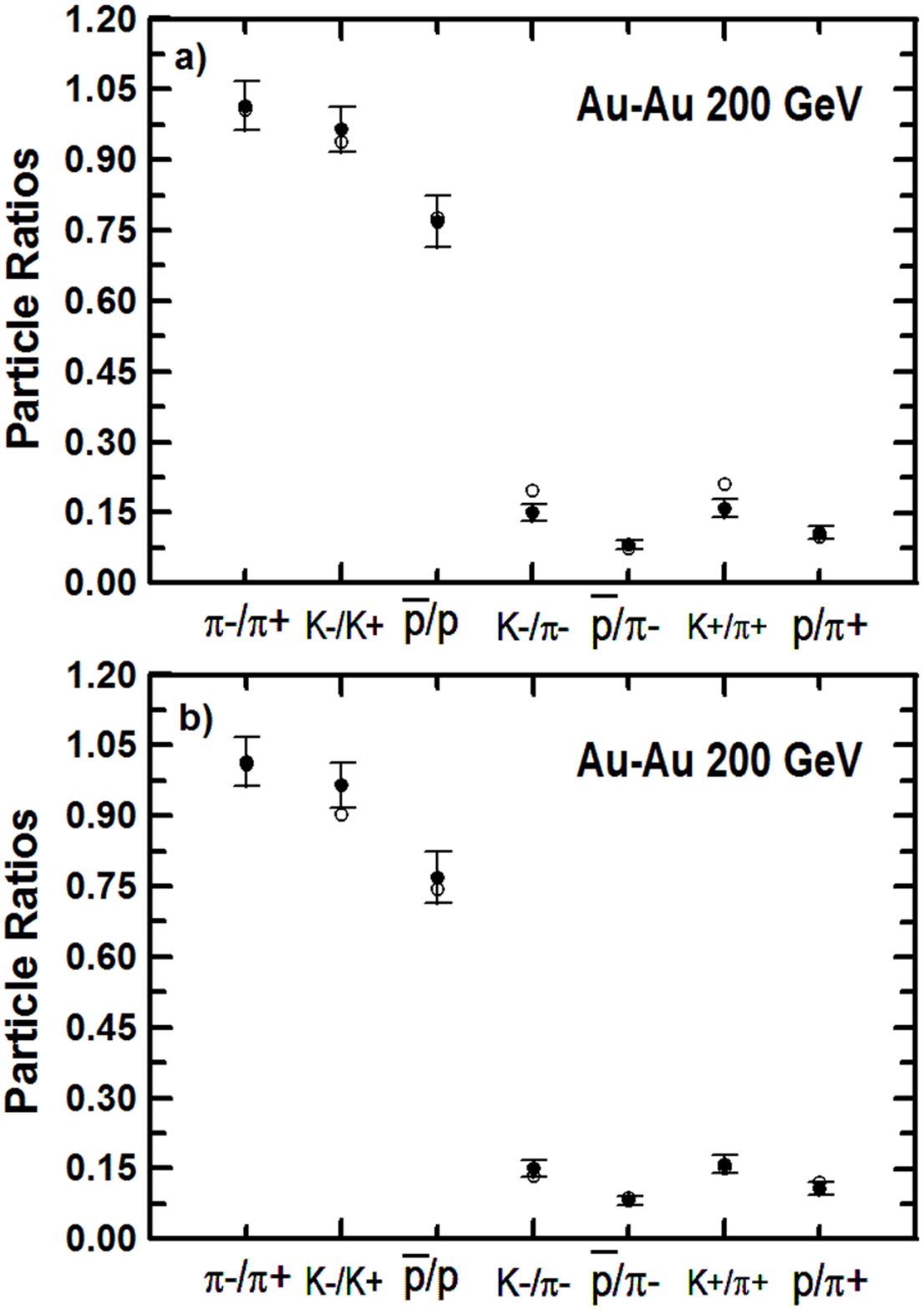}
	\caption{\label{fig3}Plots for particle ratios of Au + Au at 200 GeV fitted using the $1T$-Model in part (a), and the $2T$-Model, in (b), both with open circles. Full circles with error bars are the data taken from Ref.\cite{Abelev}.}
\end{figure}

In Figures \ref{fig2} and \ref{fig3} we show the obtained best fitting of particles ratios results corresponding to the system in previous figure compared with the data. Although the fitting seems similar in the plots the rather good improvement in the quality by using $2T$-Model can be seen comparing the $\chi_{d.o.f}^{2}$ values in previous tables.

\clearpage

\section{The collision energy dependence of the Freeze-out temperatures, baryochemical potential and entropy}

From the results presented in Tables \ref{table1} and \ref{table2} we can exhibit a sketch of the baryonic and mesonic freeze-out temperatures and baryochemical potential behavior as a function of the collision energy $\sqrt{S_{NN}}$. With the five points, corresponding to the chosen systems, we fitted the results of the thermal parameters  as function of the collision energy, $\sqrt{S_{NN}}$. For the temperatures we have used the fitting expression similar to that one considered in Ref. \cite{andronic2009}, 

\begin{equation}
T=\frac{164c}{1+av^{-b}}, \label{fitt}
\end{equation}

\noindent where $v=\sqrt{S_{NN}}$. For baryonic potential the parametrization used was \cite{andronic2009}

\begin{equation}
\mu_{b}={\frac{d}{1+ev}.} \label{fitm}
\end{equation}

The parameters values for the best fitting in Eqs.~(\ref{fitt}) and (\ref{fitm}) are showed in Tables \ref{table3a} and \ref{table3b} for both $1T$ and $2T$ models. 

\begin{figure}[h]
	\includegraphics[width=140mm]{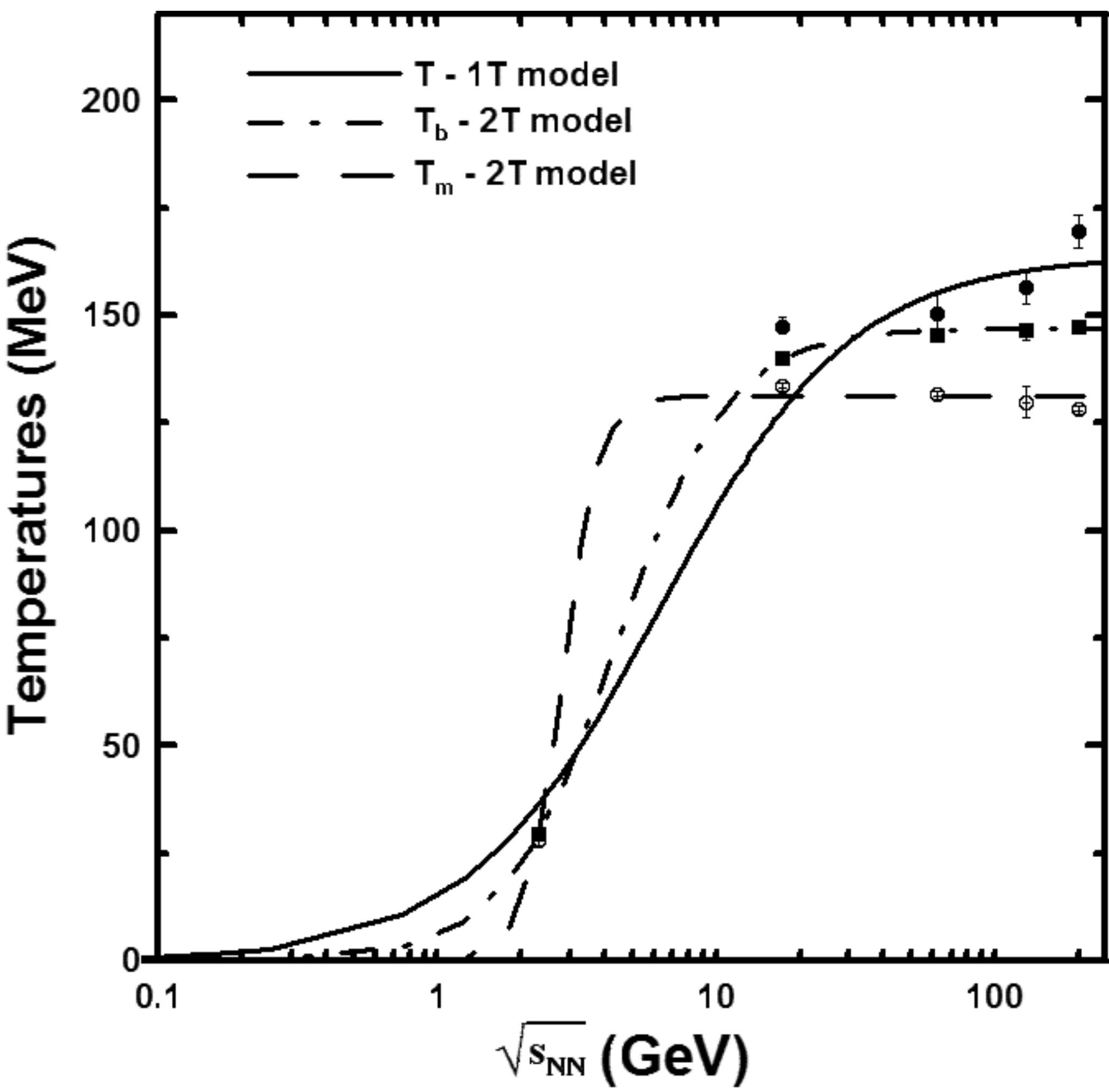}
	\caption{\label{fig4}Plots of temperature against the collision energy.
 The plot with full-line and full-circles represents the parametrization and fitted data sets for the $1T$-Model. For the $2T$-Model the plot are dashed-dotted line (parametrization) with full squares (fitted data sets) corresponding to  $T_b$ and dashed line (parametrization) with open circles (fitted data sets) to $T_m$.}
\end{figure}

\begin{figure}[h]
	\includegraphics[width=140mm]{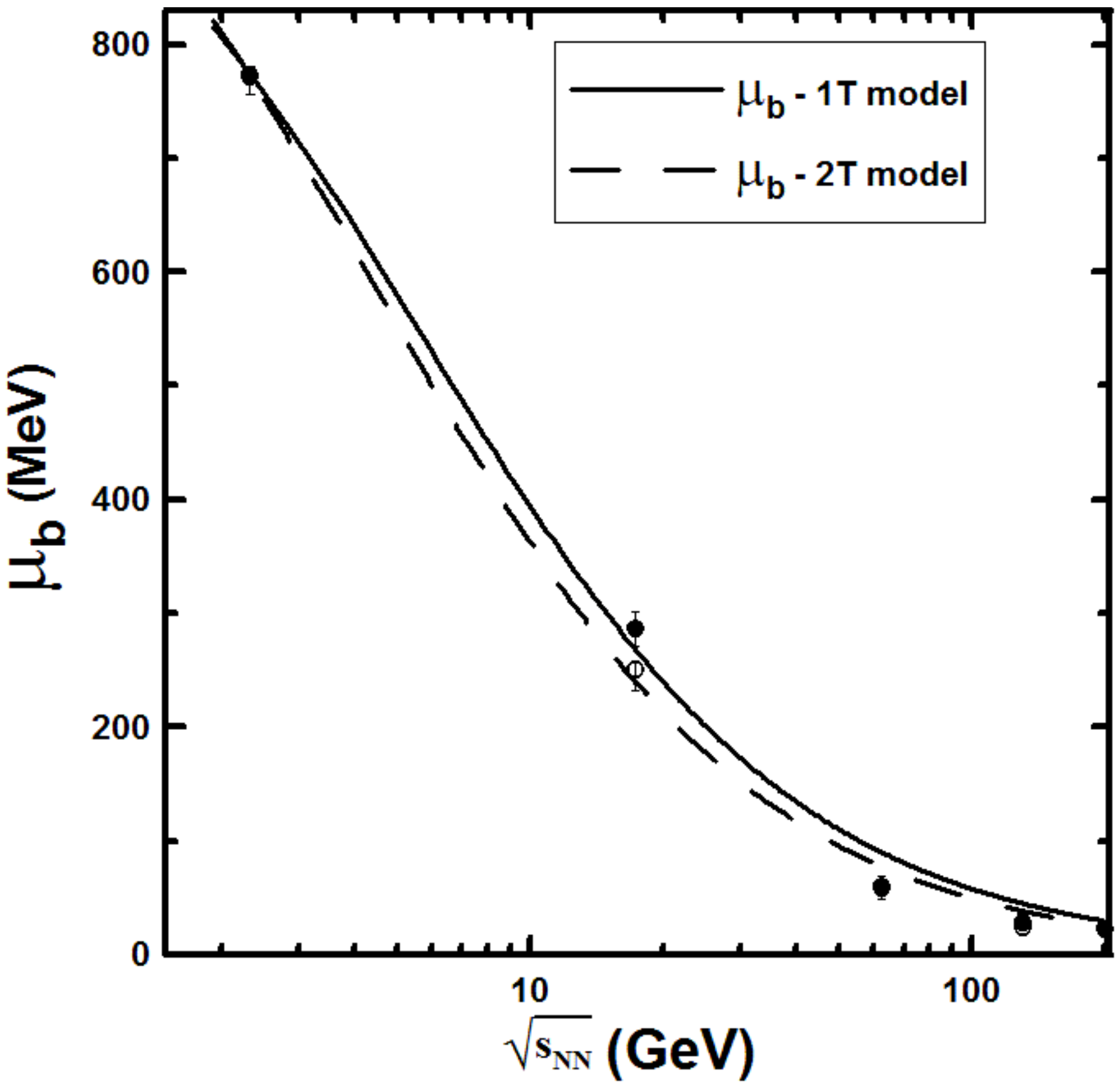}
	\caption{\label{fig5}Plots of baryonic potential against the collision energy.
 The plot with full line and open circles represents the parametrization and fitted data sets for the $1T$-Model. For the $2T$-Model
the plot with dashed line with open circles is the parametrization and fitted data set.}
\end{figure}

In Figs. \ref{fig4} and \ref{fig5} it is shown a plot of temperature and baryonic chemical potential dependence with the collision energy. As observed in many others calculations, a sudden increase is observed in the temperatures (Fig.~\ref{fig4}) in the low energy regime and a constant value reached asymptotically. Our result shows a quite defined plateau for the $2T$-Model reached quickly with the energy increasing, that is not the case for the $1T$-Model result in which it is not formed for the covered collision energy values.

In Fig.~\ref{fig5} it is shown the freeze out baryochemical potential as a function of the collision energy for the case of  $1T$ and $2T$ models. As the meson sector does not carries baryonic charge, the results for the baryochemical potential cannot differ significantly when using $(1T)$ or $(2T)$ description. A small difference is observed due to the  effect of the temperatures on the constraints coupling (Eqs.~(\ref{eq1}-\ref{qs} )) and on the feeding process in Eq.~(\ref{feed}).

\begin{table}[h,t]
\caption{\label{table3a}Parameters of the temperature fitting in Eq.~(\ref{fitt}). First line corresponds to the fit within the 1$T$-Model, the other two lines are the fittings within the 2$T$-Model.}
\begin{tabular}{|c|c|c|c|} \hline\hline
         & $a$            & $b$        & $c$  \\ \hline
 $T$ & $19.026$ & $1.739$ & $1.0$         \\ \hline
$T_{b}$ & $24.108$ & $2.145$ & $0.897$  \\
$T_{m}$ & $1115.926$ & $6.783$ & $0.800$  \\ \hline\hline
\end{tabular}
\end{table}

\begin{table}[h,t]
\caption{\label{table3b}Parameters of the chemical potential fitting in Eq.~(\ref{fitm}). First line corresponds to the fit within the 1$T$-Model, the second line is the fittings within the 2$T$-Model.}
\begin{tabular}{|c|c|c|}\hline\hline
                    & $d$ & $e$\\\hline
$\mu_{b}$ & $1094.050$ & $0.178$\\\hline
$\mu_{b}$ & $1177.078$ & $0.225$\\ \hline\hline
\end{tabular}
\end{table}

In $2T$-Model the entropy for mesonic or baryonic sectors are calculated separately as,

\begin{equation}
S_i=\frac{1}{T_i}\left(\varepsilon_i + P_i - \sum_k \mu_k n_k \right)
\end{equation}

\noindent where $i=m$(meson), $b$(baryon) specifies the sector, with $\varepsilon_i$ and $P_i$ being the corresponding energy density and pressure. The $k$ subscript refers to the particles entering in the sector composition. It is well known from conventional models that the dimensionless quantity $S/T^3$ increases (decreases) for mesons (baryons) as $\sqrt{S_{NN}}$ increases \cite{Cleymans2009}. This behavior is claimed as a change from a baryon dominance to a meson  dominance regime of the collision freeze-out, marked by the energy value at the point where $S_m/T^3$ and $S_b/T^3$ curves crosses.  From Table~\ref{table5} this change of regime is also observed in the $2T$-Model,  where can be identified a meson dominance regime around $\sqrt{S_{NN}}$ = $7$ GeV.

\begin{table}[h]
\caption{\label{table4}Mesonic and baryonic entropy densities for the $1T$-Model. The third and fourth columns are the mesonic and baryonic entropy densities for different collisional sets respectively. In the fifth column we can see the mesonic-baryonic entropy ratio. In the last two columns we have the mesonic (baryonic) entropy density over cubic mesonic (baryonic) temperatures.}
\begin{tabular}{|c|c|c|c|c|c|c|}\hline\hline
Set&$\sqrt{s_{NN}} \left(  GeV\right)$ &$S_{m}$ (fm$^{-3}$) &$S_{b}$ (fm$^{-3}$) &
$S_{m}/S_{b} $ & $S_{m}/T^{3} $  & $ S_{b}/T^{3} $ \\\hline
$C1$&$2.32$ & $0.669\times10^{-3}$ & $0.492\times10^{-2}$ & $0.136$ & $0.198$ & $0.146$ \\\hline
$C2$&$17.3$ & $1.940$ & $0.839$ & $2.310$ & $4.652$& $2.013$ \\\hline
$C3$&$62.4$ & $1.994$ & $0.445$ & $4.486$ & $4.495$ & $1.002$ \\\hline
$C4$&$130$ & $2.368$ & $0.581$ & $4.077$ & $4.747$ & $1.164$ \\\hline
$C5$&$200$ & $3.230$ & $1.008$ & $3.205$ & $5.096$ & $1.590$\\ \hline\hline
\end{tabular}
\end{table}

\begin{table}[h]
\caption{\label{table5}Same as Table \ref{table4}, this time for the $2T$-Model.}
\begin{tabular}{|c|c|c|c|c|c|c|}\hline\hline
Set&$\sqrt{s_{NN}} \left(  GeV\right)$ &$S_{m}  $ (fm$^{-3}$) &$S_{b}  $ (fm$^{-3}$) &
$S_{m}/S_{b} $ & $S_{m}/T_{m}^{3} $  & $ S_{b}/T_{b}^{3} $ \\\hline
$C1$&$2.32$ & $0.471\times10^{-3}$ & $0.479\times10^{-2}$ & $0.0984$ & $0.167$ & $1.418$ \\\hline
$C2$&$17.3$ & $1.226$ & $0.524$ & $2.340$ & $3.971$& $1.465$ \\\hline
$C3$&$62.4$ & $1.086$ & $0.370$ & $2.937$ & $3.669$ & $0.922$ \\\hline
$C4$&$130$ & $1.023$ & $0.379$ & $2.702$ & $3.600$ & $0.929$ \\\hline
$C5$&$200$ & $0.974$ & $0.408$ & $2.384$ & $3.568$ & $0.981$\\ \hline\hline
\end{tabular}
\end{table}

\section{Conclusions}

In this work we have suggested a non-equilibrium scenario for the thermal freeze-out in heavy-ion collisions. Distinct temperatures are introduced to handle with thermal densities of baryons and mesons. Conceptually, these temperatures are associated to the thermal state of the vacuum from where particles are produced. In our calculation (Figures \ref{fig1}-\ref{fig3}) we have shown that within the two-temperatures model we can obtain a rather good fits for the ratios of particles population in a wide region of parameters values (Figures \ref{fig3}-\ref{fig4}). A similar situation was also obtained for other analyzed systems.  
Regarding the fitting of particle ratios to the data, we did not privilege any one of the mesonic or baryonic sectors. Our results from the two-temperatures model  revealed the best fitting when the baryonic temperature is higher than the mesonic temperature. Although previous investigations about different temperatures for meson and baryon sectors in thermal models have been done in regard to Hagedorn's states \cite{plb2004}, we remark that in our calculation the attributed temperatures are conceptuality differentiated from Hagedorn's temperatures.

Concerning to the validation of the model, we are aware that further explorations with a comprehensive data sets for particles rate are necessary, as well as the analysis of other observables extracted from the experiments. In general, for a given reaction, the set of detected particle population includes only a few types of particle ratios and it is a hard task to obtain a confident model matching. In addition, it should be noted that it is usual to find in the literature different data sets coming from the same experiment, since they are constrained to different data filters or selection criteria. Even so, the performed calculation (see results in Figures \ref{fig1}-\ref{fig5}) have shown that using the two-temperatures thermal model the quality of the fittings for the ratios of particles population were improved in a wide region of the parameters values. This opens room enough to accommodate predictions of observable not explored here, such as strangeness production, transverse distribution of momentum, total particle-antiparticles ratios, and others. 

Finally, we remark that the effect of using the two-temperatures model on other collision systems can be enhanced since the  conservation laws given by Eqs.~(\ref{qb})-(\ref{qs}) play a central role in the calculation.  

\begin{acknowledgments}
Three of us (L.P.G.A, S.B.D and A.D.J) want to thanks CNPq for finantial support.
\end{acknowledgments}


\begin{thebibliography}{99}                                                                                               %

\bibitem {kraus2009a}I. Kraus, J. Cleymans, H. Oeschler, K. Redlich, and S. Wheaton, Prog. Part. Nucl. Phys. \textbf{62},  538 (2009).

\bibitem {kraus2009}I. Kraus, J. Cleymans, H. Oeschler, and K. Redlich, Phys. Rev. C \textbf{79}, 014901 (2009).

\bibitem {kraus2007}I. Kraus, J. Cleymans, H. Oeschler, K. Redlich, and S. Wheaton, Phys. Rev. C \textbf{76}, 014901 (2009).

\bibitem {cleymans1994r}J. Cleymans, K. Redlich, H. Satz, and E. Suhonen, Z. Phys. C \textbf{58}, 347 (1994). 

\bibitem {rafelski1991r}J. Rafelski, Phys. Lett. B \textbf{262}, 333 (1991).

\bibitem {becattini1996r}F. Becattini,  Z. Phys. C \textbf{69}, 485 (1996).

\bibitem {becattini1998r}F. Becattini, M. Gazdzicki, and J. Sollfrank,  EPJC \textbf{5}, 143 (1998).

\bibitem {becattini2009}F. Becattini,  arXiv:0901.3643v.

\bibitem {rischke1991r}D. H. Rischke, M. I. Gorenstein, H. St\"ocker, and W. Greiner,  Z. Phys. C \textbf{51}, 485 (1991).

\bibitem {yen1999r}G. D. Yen, and M. I. Gorenstein, Phys. Rev. C \textbf{59}, 2788 (1999).

\bibitem {braum1999}P. Braun-Munzinger, I. Heppe, and J. Stachel,  Phys. Lett. B  \textbf{465}, 15 (1999).

\bibitem {Cleymans1999}J. Cleymans, H. Oeschler, and K. Redlich, Phys. Rev. C \textbf{59}, 1663 (1999).

\bibitem {andronic2009}A. Andronic, P. Braun-Munzinger, and J. Stachel, Phys. Lett. B \textbf{673},  142 (2009).
 
\bibitem {Cleymans2009}J. Cleymans \emph{et al.},  APPB Proc. Suppl. \textbf{3}, 495 (2009).

\bibitem {hirsh2010} L. R. Hirsch, Ph.D. Thesis, Universidade Federal Fluminense, 2010.

\bibitem {hirsh2010b}L. R. Hirsch, A. Delfino, and M. Chiapparini,  Nucl. Phys. B Proc. Suppl. \textbf{199}, 297 (2010).

\bibitem {bass2000}S. A. Bass, and A. Dumitru, Phys. Rev. C \textbf{61}, 064909 (2000).

\bibitem {song2011} H. Song, S. A. Bass, and U. Heinz,  Phys. Rev. C \textbf{83}, 024912 (2011).

\bibitem {hirano2002}T. Hirano, and K. Tsuda, Phys. Rev. C \textbf{66}, 054905 (2002).

\bibitem {beibe1992}H. Bebie, P. Gerber, J.L. Goity, and H. Leutwyler, Nucl. Phys. B \textbf{378}, 95 (1992).

\bibitem {kolb2003}P. F. Kolb, and R. Rapp, Phys. Rev. C \textbf{67}, 044903 (2003).

\bibitem {huovinen2008}P. Huovinen, EPJA \textbf{37}, 121 (2008).

\bibitem {Muronga} A. Muronga,  Phys. Rev. Lett. \textbf{88}, 062302 (2002). 

\bibitem {noronha11} R. Rapp, and E. V. Shuryak,  Phys. Rev. Lett. \textbf{86}, 2980 (2001).


\bibitem {noronha12} C. Greiner,  AIP Conf. Proc. \textbf{644}, 337 (2003); C. Greiner and S. Leupold,   J. Phys. G: Nucl. Part. Phys. \textbf{27}, L95 (2001).

\bibitem {noronha13} P. Koch, B. Muller, and J. Rafelski,  Phys. Rep. \textbf{142}, 167 (1986).

\bibitem {noronha14} J. I. Kapusta, and I. Shovkovy,  Phys. Rev. C \textbf{68}, 014901 (2003);
J. I. Kapusta,   J. Phys. G: Nucl. Part. Phys. \textbf{30}, S351 (2004).

\bibitem{plb2004} W. Broniowski, and W. Florkowski, Phys. Lett. B \textbf{490}, 223 (2004).

\bibitem{cohen2011} T. D. Cohen, and V. Krejcirik, J. Phys. G: Nucl. Part. Phys. \textbf{39}, 055001 (2012). 

\bibitem {noronha15} P. Huovinen, and J. I. Kapusta,  Phys. Rev. C \textbf{69}, 014902 (2004).

\bibitem {noronha} J. Noronha-Hostler, H. Ahmad, J. Noronha, and C. Greiner, Phys. Rev. C \textbf{82}, 024913 (2010). 

\bibitem {lu2002b} Z. D. Lu, B. H. Sa, A. Faessler, C. Fuchs, and E. E. Zabrodin, Phys. Rev. C \textbf{66}, 044905 (2002).

\bibitem {choi2011}S. Choi, and K. S. Lee,  Phys. Rev. C \textbf{84}, 064905 (2011); Phys. Rev. C \textbf{84}, 069901 (2011).

\bibitem{gge} R. Balian, \emph{From Microphysics to Macrophysics: Methods and Applications of Statistical Physics} (Springer, Berlin, 1991), Vol. 2.

\bibitem {Fermi50} E. Fermi, Prog. Theor. Phys. \textbf{5}, 570 (1950).

\bibitem {Heis49} W. Heisenberg, Nature \textbf{164}, 65 (1949).

\bibitem {Abelev} B. I. Abelev \emph{et al.}, Phys. Rev. C \textbf{79}, 034909 (2009).

\end{thebibliography}
\end{document}